\documentclass[
    aps,
    pre,
    amsmath,
    amssymb,
    reprint,
    superscriptaddress,
    preprintnumbers,
]{revtex4-2}

\usepackage[utf8]{inputenc}
\usepackage[T1]{fontenc}
\usepackage[english]{babel}

\usepackage{amsthm}
\usepackage{physics}
\usepackage{bm}

\usepackage{graphicx}
\usepackage{tikz}
\usepackage{pgf}
\usepackage{booktabs}
\usepackage{dcolumn}

\usepackage[hidelinks]{hyperref}

\begin{document}

\title{Diagnosing quantum reservoirs at scale based on expressivity and coverage}

\author{Laia Domingo}
\email{ldomingo@cvc.uab.cat}
\thanks{ORCID:
\href{https://orcid.org/0000-0003-3535-3538}
{0000-0003-3535-3538}}
\affiliation{Centre de Visió per Computador (CVC), Barcelona, Spain}
\affiliation{Universitat Autònoma de Barcelona (UAB), Barcelona, Spain}

\author{Oriol Balló-Gimbernat}
\affiliation{Centre de Visió per Computador (CVC), Barcelona, Spain}
\affiliation{Universitat Autònoma de Barcelona (UAB), Barcelona, Spain}
\affiliation{Eurecat, Centre Tecnològic de Catalunya, Barcelona, Spain}

\author{Fernando Vilariño}
\affiliation{Centre de Visió per Computador (CVC), Barcelona, Spain}
\affiliation{Universitat Autònoma de Barcelona (UAB), Barcelona, Spain}


\begin{abstract}
Quantum reservoirs offer a hardware-friendly route to quantum machine
learning, replacing trainable circuits with fixed random dynamics and a
classical readout. Because the reservoir is not optimized, performance
depends entirely on the choice of reservoir family, yet existing diagnostics
demand resources that grow exponentially with system size. We introduce a
scalable, hardware-agnostic framework built on two complementary quantities.
The first is a task-independent order-statistics (ORS) expressivity score,
which compares only the largest output probabilities of a reservoir ensemble
against an analytical Haar baseline. It never reconstructs the full output
distribution, is cost-independent of Hilbert-space dimension, and admits a
closed-form depolarizing-noise correction, making it directly usable on
hardware. The second is the task-dependent effective rank
$R_{\mathrm{eff}}$ of the feature matrix, which measures how much
input-dependent information reaches the readout. We validate the ORS score
against established complexity diagnostics and confirm it remains informative
under simulated noise and on IBM quantum hardware. Across synthetic and real
quantum extreme learning machine and quantum reservoir computing benchmarks,
ORS captures the intrinsic expressivity hierarchy of reservoir families while
$R_{\mathrm{eff}}$ determines when that expressivity becomes usable
predictive information.
\end{abstract}

\maketitle

\section{Introduction}

Quantum reservoirs (QRs) are among the most hardware-friendly approaches to quantum machine learning on near-term devices. Inherited from classical reservoir computing~\cite{jaeger2004harnessing,
maass2004computational, tanaka2019recent, nakajima2020physical}, their core idea is to process data with the dynamics of a fixed quantum system. Each data point is embedded into the reservoir state, evolved through the quantum dynamics, and represented by measured observables. These observables form a feature vector that is passed to a classical readout, usually a linear regression. Since the quantum system itself is not trained, QRs avoid the optimization pathologies of variational quantum algorithms, most notably barren plateaus~\cite{mcclean2018barren, cerezo2021cost,
larocca2024review, anschuetz2022traps, thanasilp2023subtleties}, while still accessing nonlinear quantum features in a Hilbert space that grows exponentially with the number of qubits~\cite{fujii2017harnessing, mujal2021opportunities}. In some regimes, noise can even become a useful computational resource rather than only a source of degradation~\cite{DomingoSciRep2023NoiseQRC, sannia2024engineered}.

This mechanism appears in two closely related settings. In quantum reservoir computing (QRC), the reservoir processes temporal signals and retains memory of past inputs, making it suitable for time-series tasks~\cite{chen2020temporal, kobayashi2024feedback,
hou2026experimental}. In quantum extreme learning machines (QELMs), inputs are processed independently, which is suitable for static regression and classification~\cite{xiong2024fundamental}. Both paradigms have been explored in digital and analog platforms~\cite{senanian2024microwave, nokkala2021gaussian,
govia2021single}, with applications ranging from quantum chemistry~\cite{domingo2022optimal, quantumchemQRC} to hybrid quantum-classical neural networks~\cite{BindingAffinities}.

Because the reservoir is fixed, performance depends critically on the choice of its random dynamics. A reservoir that is too simple cannot generate a useful feature map, while a highly complex reservoir may still fail if the measured features do not expose enough task-relevant information to the readout. Several diagnostics have been proposed to characterize reservoir quality, including level-spacing statistics~\cite{martinezpena2021dynamical}, majorization~\cite{latorre2002majorization,
vallejos2021majorization}, entanglement-spectrum statistics~\cite{shaffer2014entanglement}, Krylov complexity~\cite{domingo2024krylov}, and expressivity measures based on the Kullback--Leibler (KL) divergence to the Haar fidelity distribution~\cite{expressibility}. However, these tools are either tied to specific physical models or require resources that scale poorly with system size. Full-state, full-unitary, or full-distribution diagnostics rapidly become impractical, and fidelity-based KL estimators become unreliable as the Haar fidelity distribution concentrates exponentially with the number of qubits.
In this work, we introduce a scalable, hardware-compatible framework for diagnosing quantum reservoirs along two complementary axes: \emph{expressivity} and \emph{coverage}. Our contribution is threefold. First, we adapt the order-statistics (ORS) score, recently introduced for simulation-free fidelity estimation of quantum circuits~\cite{micklitz2025ors}, into a task-independent expressivity diagnostic for reservoir ensembles. Rather than reconstructing the full output distribution, the method compares only the largest output probabilities with an analytical Haar order-statistics baseline. The number of retained probabilities can therefore be fixed independently of Hilbert-space dimension, allowing the diagnostic to remain tractable as the number of qubits increases. The resulting score is applicable to arbitrary reservoir families and admits a closed-form correction for depolarizing noise, making it directly usable with finite-shot and hardware data. Second, we introduce a multi-basis extension that evaluates the ORS gap over random local Pauli measurement bases. Because Haar-random states are basis invariant, this extension distinguishes genuinely Haar-like reservoir ensembles from families whose apparent randomness is restricted to a privileged measurement basis.

Third, we combine this task-independent notion of expressivity with a task-dependent measure of \emph{coverage}, given by the effective rank $R_{\mathrm{eff}}$ of the feature matrix. While ORS characterizes the intrinsic complexity of the reservoir ensemble, $R_{\mathrm{eff}}$ quantifies how many independent, input-dependent feature directions are actually exposed to the classical readout. This distinction is essential: expressive reservoir dynamics improve performance only when they generate a sufficiently rich feature matrix, which can depend on other factors like the data encoding or observables measured \cite{xiong2024fundamental}. Moreover, $R_{\mathrm{eff}}$ can decrease because of structural symmetries of the reservoir, as in commuting instantaneous quantum polynomial (IQP) families, or because of exponential concentration, where observables become increasingly insensitive to the input as system size grows~\cite{xiong2024fundamental, sannia2025concentration,
thanasilp2024kernels}.

We validate this two-axis picture across complementary complexity diagnostics, noise regimes, hardware experiments, and learning tasks. In the small-system regime, ORS reproduces the expressivity transition identified independently by KL divergence, spectral chaos indicators, and Krylov complexity. Under simulated depolarizing noise and execution on IBM quantum hardware, the noise-corrected ORS gap continues to separate expressive from structurally restricted reservoir families. Across synthetic and real QELM and QRC benchmarks, multi-basis ORS captures the intrinsic expressivity hierarchy of the reservoirs, while $R_{\mathrm{eff}}$ identifies when this expressivity is converted into useful predictive features. Together, ORS and $R_{\mathrm{eff}}$ provide a scalable and hardware-compatible framework for diagnosing QRs across architectures, tasks, and noise regimes.


\section{Methods}
\label{sec:methods}
\subsection{Quantum reservoirs: QRC and QELM}
\label{sec:methods-qr}

A QR encodes an input into a quantum state, evolves it under a fixed random reservoir unitary $U$ sampled once from a chosen family, and reads out the local observables that feed a classical linear regression. In the QRC setting, the inputs form a temporal sequence $\{x_t\}$ and the quantum state is updated recursively. The $n$ qubits are split into three roles: \emph{input} qubits, which encode the data at each step; \emph{measurement} qubits (which usually include the input qubits), whose local observables are read out; and
\emph{hidden} qubits, which are never measured and retain memory of past inputs. At time $t$, the input qubits are prepared in the data-encoding state $\ket{\psi_t}$, combined with the retained state of the remaining qubits, evolved under the unitary $U$, and measured:
\begin{eqnarray}
    &\rho_t = U\,\big(\ket{\psi_t}\!\bra{\psi_t}\otimes
    \operatorname{Tr}_{n_i}[\rho_{t-1}]\big)\,U^\dagger, \\ \nonumber
    &\langle O_j\rangle_t = \operatorname{Tr}[O_j\,\rho_t],
\end{eqnarray}
where $\operatorname{Tr}_{n_i}[\cdot]$ traces out the input qubits before the next data point is injected, and $O_j$ is an observable of the system. The measured features $\mathbf{h}_t = (\langle O_1\rangle_t, \langle O_2\rangle_t, \ldots \langle O_k\rangle_t)$ feed the readout layer, which is usually a classical linear regression, such that $\hat{y}_{t+1} = W\mathbf{h}_t$.

In the QELM setting, the recurrent feedback loop is removed and each input sample is processed independently. For each input $x$, the corresponding encoding state $\ket{\psi(x)}$ is prepared, evolved once under the fixed unitary $U$, and measured. The resulting feature vector is
\begin{equation}
    \mathbf{h}(x)=\big(\langle O_1\rangle_x,\langle O_2\rangle_x,\ldots,\langle O_k\rangle_x\big),
\end{equation}
and the prediction is obtained as $\hat{y}=W\mathbf{h}(x)$. QELMs therefore implement a fixed nonlinear feature map $x\mapsto \mathbf{h}(x)$, making them naturally suited to static regression and classification tasks~\cite{xiong2024fundamental}.

\subsection{Expressivity: the order-statistics score}
\label{sec:methods-ors}
We QR expressivity by repurposing the ORS score  introduced in Ref.~\cite{micklitz2025ors}. Originally developed for simulation-free fidelity estimation of large quantum circuits, ORS uses the statistics of high-probability output bitstrings to assess whether a circuit output is consistent with Haar-random behavior. Here, we adapt this idea to QRs: instead of estimating the fidelity of a single circuit output, we use the order statistics of output probabilities to diagnose how close a reservoir ensemble is to Haar-random measurement statistics.

The key advantage of this approach is that it does not require reconstructing the full output distribution. For each reservoir instance, we only retain the $K$ most likely computational-basis outcomes, equivalently the most frequently observed bitstrings in a finite-shot experiment. If the circuit instance produces probabilities $\{p_j\}_{j=1}^{D}$ over the $D=2^n$ basis states, we denote their ordered values by
\begin{equation*}
    p_{(1)} \geq p_{(2)} \geq \cdots \geq p_{(D)} .
\end{equation*}
For a Haar-random state, the distribution of these ordered probabilities is known analytically. In particular, if $x$ denotes a possible value of the $k$-th largest probability $p_{(k)}$, then $P_k(x)$ is the probability density for observing $p_{(k)}=x$ under the Haar ensemble. In the large-$D$ limit, this density takes the asymptotic form
\begin{equation}
    P_k(x) \propto e^{-k a}\bigl(1-e^{-a}\bigr)^{D-k},
\quad a=(D-2)x,
\end{equation}
up to a normalization factor given in Appendix~\ref{app:ors}. Since only the first $K$ ordered probabilities are evaluated, the number of retained outcomes can be fixed independently of the Hilbert-space dimension. In practice, we take $K$ to be small, typically scaling at most linearly with the number of qubits, $K\sim\mathcal{O}(n)$.

Given an ensemble of $M$ reservoir instances, we evaluate the Haar log-likelihood of the observed top-$K$ ranks,
\begin{equation}
\overline{\Lambda}
= \frac{1}{M}\sum_{m=1}^{M}\sum_{k=1}^{K}
\frac{1}{k}\log P_k\bigl(p_{(k)|m}),
\label{eq:lambda-score}
\end{equation}
where $p_{(k)|m}$ is the $k$-th largest output probability of the $m$-th reservoir instance. The harmonic weights $1/k$ emphasize the highest-probability outcomes, which carry the strongest order-statistical signal. The Haar reference value, denoted $\Lambda_{\rm Haar}$, is obtained by evaluating the same quantity analytically under the Haar order-statistics distribution, as detailed in Appendix~\ref{app:ors}. We then define the ORS expressivity diagnostic as the signed gap
\begin{equation}
G_{\rm ORS} = \overline{\Lambda} - \Lambda_{\rm Haar}.
\label{eq:ors-gap}
\end{equation}
Reservoir ensembles whose top-ranked output probabilities follow Haar order statistics have $G_{\rm ORS}\simeq 0$, while deviations from zero indicate departures from Haar-random behavior. This provides a scalable diagnostic of QR expressivity that is defined for arbitrary reservoir families, depends only on a small number of measured bitstrings, and can be applied directly to noisy data.

\subsection{Hardware-aware expressivity under noise}
\label{sec:methods-noise}

The ORS diagnostic has a natural extension to depolarizing noise, already used in Ref.~\cite{micklitz2025ors}, which makes it suitable for hardware-aware expressivity estimation. In the ideal setting, ORS compares the top-ranked output probabilities of a circuit ensemble with the Haar order-statistics distribution. On noisy hardware, however, even a Haar-typical circuit is not expected to reproduce the ideal Haar statistics directly: noise contracts the output distribution toward the uniform distribution. The appropriate reference is therefore not the ideal Haar ensemble, but a Haar-random ensemble affected by the same effective noise level. We model this effect with a single-parameter \textit{global} depolarizing channel. If $f$ denotes the effective circuit fidelity, each ideal output probability is mapped to
\begin{equation}
p_j^{(f)}
=
f p_j + \frac{1-f}{D}.
\label{eq:depol-probability}
\end{equation}
For $f>0$, this affine transformation preserves the ordering of the probabilities. Thus, if $x$ is a noisy ranked probability, the corresponding ideal probability variable is
\begin{equation}
x_f
=
\frac{x-(1-f)/D}{f}.
\label{eq:inverse-depol-map}
\end{equation}
The Haar order-statistics density is then transformed as
\begin{equation}
P_k^{(f)}(x)
=
\frac{1}{f} P_k(x_f),
\label{eq:noisy-ors-density}
\end{equation}
where the prefactor $1/f$ is the Jacobian of the affine change of variables. The noisy ORS score is obtained by replacing the ideal Haar density $P_k$ with $P_k^{(f)}$ in the log-likelihood,
\begin{equation}
\overline{\Lambda}^{(f)}
=
\frac{1}{M}
\sum_{m=1}^{M}
\sum_{k=1}^{K}
\frac{1}{k}
\log P_k^{(f)}
\left(
p_{(k)|m}^{(f)}
\right).
\label{eq:noisy-lambda-score}
\end{equation}
The corresponding Haar reference is shifted analytically by the same Jacobian term:
\begin{equation}
\Lambda_{\mathrm{Haar}}^{(f)}
=
\Lambda_{\mathrm{Haar}}
-
\sum_{k=1}^{K}\frac{1}{k} \log f.
\label{eq:noisy-haar-baseline}
\end{equation}
As expected, the noiseless baseline is recovered when $f\to 1$. A key consequence of this correction is that, under an exact global depolarizing model, the corrected ORS gap is invariant under the fidelity parameter. For fixed reservoir instances and exact probabilities, Eq.~\eqref{eq:noisy-ors-density} shifts both the reservoir score and the Haar reference by the same amount, so that
\begin{equation}
G_{\mathrm{ORS}}^{(f)} \equiv \overline{\Lambda}^{(f)}
-
\Lambda_{\mathrm{Haar}}^{(f)}
=
\overline{\Lambda}
-
\Lambda_{\mathrm{Haar}} \equiv G_{\mathrm{ORS}}.
\label{eq:noisy-gap-invariance}
\end{equation}
This invariance is important because it allows ORS gaps obtained at different effective fidelities to be compared on the same scale, provided that the dominant noise can be approximated by a global depolarizing channel. In this sense, the correction separates the intrinsic deviation of a reservoir family from Haar-random behavior from the trivial loss of contrast caused by depolarization.

We use this correction in two complementary settings. First, in simulation, we apply controlled depolarizing noise with known values of $f$ and verify Eq.~\eqref{eq:noisy-gap-invariance} by evaluating the same reservoir instances across fidelities. Second, in experiments on IBM quantum hardware, we estimate an effective fidelity for each transpiled circuit from backend calibration data. Let $C_{\mathrm{tr}}$ be the transpiled circuit executed on backend $\mathcal{Q}$, and let $B_q$ denote the physical block of qubits used to compute the ORS score. We estimate the total block error by 
\begin{equation}
E_{B_q}(C_{\mathrm{tr}},\mathcal{Q})
=
E_{B_q}^{1q}
+
E_{B_q}^{2q}
+
E_{B_q}^{\mathrm{ro}},
\label{eq:block-error-decomposition}
\end{equation}
where $E_{B_q}^{1q}$ collects the errors of one-qubit gates, $E_{B_q}^{2q}$ those of two-qubit gates, and $E_{B_q}^{\mathrm{ro}}$ the readout errors of the measured qubits. The corresponding effective hardware fidelity is then defined as
\begin{equation}
f_{B_q}(C_{\mathrm{tr}},\mathcal{Q})
=
\exp\left[
-
E_{B_q}(C_{\mathrm{tr}},\mathcal{Q})
\right].
\label{eq:block-fidelity}
\end{equation}
When the transpiled circuits in an ensemble have different gate contents, this fidelity is computed separately for each circuit instance and used in Eq.~\eqref{eq:noisy-lambda-score}. Although real device noise is not exactly global depolarizing, this hardware-aware correction provides an effective way to compare reservoir families executed at different noise levels.
\subsection{Multi-basis order-statistics extension}
\label{sec:methods-multi-basis}

The order-statistics score defined above probes the output distribution in a chosen measurement basis. In the simplest implementation, this is the computational basis, which we denote by $Z$. However, Haar-random states are basis invariant: if an ensemble is genuinely Haar-like, then its measurement statistics should follow the same Haar order-statistics distribution after any fixed change of basis. A single-basis ORS score should therefore be interpreted as a one-basis approximation to a basis-averaged expressivity diagnostic. To make this explicit, we introduce a finite set of measurement bases $\mathcal{B}$. Each basis $b \in \mathcal{B}$ is implemented by local basis rotations before computational-basis measurement. We write
\begin{equation}
    V_b=\bigotimes_{i=1}^{n} V_{b_i},
\
V_Z=I,
\
V_X=H,
\
V_Y=H S^\dagger,
\end{equation}
For a reservoir unitary $U$ and input state
$|\psi\rangle$, the probabilities measured in basis $b$ are
\[
p_z^{(b)}
=
\left|
\langle z|V_b U|\psi\rangle
\right|^2 .
\]
For each basis $b$, we sort the probabilities as
$p_{(1)}^{(b)} \geq \cdots \geq p_{(D)}^{(b)}$ and compute
\[
\overline{\Lambda}^{(b)}
=
\frac{1}{M}
\sum_{m=1}^{M}
\sum_{k=1}^{K}
\frac{1}{k}
\log P_k
\left(
p_{(k)|m}^{(b)}
\right).
\]
Since the Haar ensemble is invariant under fixed basis changes, the Haar reference $\Lambda_{\mathrm{Haar}}$ is the same for all $b$. We define the multi-basis ORS gap as

\begin{equation}
    G_{\mathrm{mb}}(\mathcal{B})
=
\frac{1}{|\mathcal{B}|}
\sum_{b\in\mathcal{B}}
\overline{\Lambda}^{(b)}
-
\Lambda_{\mathrm{Haar}} .
\label{eq:multi-basis-ors}
\end{equation}
This quantity measures the average deviation from Haar order statistics across the sampled measurement bases. The same noise correction introduced in Eqs.~\eqref{eq:noisy-ors-density} and~\eqref{eq:noisy-haar-baseline} can be applied basis by basis to Eq.~\eqref{eq:multi-basis-ors}. In the experiments below, $\mathcal{B}$ consists of $B$ random local Pauli bases, with each qubit basis sampled independently from $\{X,Y,Z\}$. This provides a stricter and more robust expressivity test than evaluating ORS in a single measurement basis.

\subsection{Coverage: the effective rank}
\label{sec:methods-reff}

Expressivity diagnostics quantify intrinsic properties of a reservoir family, such as how close its measurement statistics are to Haar-random behavior~\cite{vallejos2021majorization}, how strongly it entangles~\cite{shaffer2014entanglement}, or how complex its dynamics are~\cite{domingo2024krylov}. ORS belongs to this class: it diagnoses the reservoir ensemble independently of the downstream task. However, learning performance also depends on whether the chosen inputs, data encoding, observables, and readout expose this expressivity as useful features.

We therefore complement ORS with a task-dependent diagnostic of \emph{coverage}, defined from the feature matrix. Given training inputs $\{x_i\}_{i=1}^{N_{\rm train}}$, let $\mathbf{h}(x_i)$ be the vector of measured observables for input $x_i$, and let $H$ be the matrix whose rows are these feature vectors. We denote the column-centred matrix by $H_{\rm c}=H-\overline{H}$. If $\{\sigma_i\}$ are the singular values of $H_{\rm c}$, we define the effective rank as the participation ratio
\begin{equation}
R_{\rm eff}(H_{\rm c})
=
\frac{\big(\sum_i \sigma_i\big)^2}{\sum_i \sigma_i^2}.
\label{eq:reff}
\end{equation}
This quantity is a smooth measure of rank: instead of counting singular values above an arbitrary threshold, it estimates how many independent feature directions carry appreciable weight. A large $R_{\rm eff}$ indicates that the readout has access to many independent input-dependent directions, whereas a small $R_{\rm eff}$ signals that the reservoir features effectively lie in a low-dimensional subspace.

This distinction is especially important in the presence of exponential concentration. In that regime, observable expectation values become increasingly insensitive to the input as the number of qubits grows, so that $\mathrm{Var}_x(\langle O_j\rangle_x)$ can decay exponentially with system size~\cite{xiong2024fundamental, sannia2025concentration}. Such concentration may be an unavoidable property of a given encoding, observable set, or reservoir family. However, $R_{\rm eff}$ provides a practical diagnostic of whether its effects are already relevant at the system size being used. If concentration is strong enough at a given $n$, the rows of $H$ become nearly indistinguishable, the singular values of $H_{\rm c}$ collapse, and $R_{\rm eff}$ decreases. In this case, even an intrinsically expressive reservoir may expose only a small number of usable feature directions to the readout.

The same interpretation extends to finite-shot experiments. A feature direction is useful only if its input-dependent variation is resolvable above measurement noise. For an observable estimated with $S$ shots, the shot-noise variance is of order $1/S$ for bounded observables. Thus, at the level of individual features, a useful signal requires
\begin{equation}
S \, \mathrm{Var}_x(\langle O_j\rangle_x) \gg 1 .
\label{eq:shot-resolvability}
\end{equation}
If this condition is not satisfied, the corresponding input-dependent variations are hidden by finite-shot fluctuations and do not contribute reliably to learning, even if they are present in the exact expectation values. Therefore, $R_{\rm eff}$ should be interpreted together with the system size $n$ and the shot budget $S$: it diagnoses whether the feature matrix available in the actual experimental regime contains enough resolvable, independent directions for the readout.

\subsection{Families of quantum reservoirs}
\label{sec:methods-families}

We consider QR families with different levels of structural complexity, using Haar-random unitaries as the maximally unstructured reference. The first set consists of random circuits generated by sequences of gates drawn from fixed gate sets. The tunable parameter is the circuit depth. We consider
\begin{align*}
G_1 &= \{\mathrm{CNOT}, H, X\}, \\
G_2 &= \{\mathrm{CNOT}, H, S\}, \\
G_3 &= \{\mathrm{CNOT}, H, T\}.
\end{align*}
The family $G_2$ generates Clifford circuits, while $G_1$ generates a subgroup of the Clifford group~\cite{gottesman1998heisenberg, clark2008generalized}. Both are therefore non-universal and classically simulable~\cite{vandennest2010classical, jozsa2014classical, koh2017further}. By contrast, the inclusion of the non-Clifford $T$ gate makes $G_3$ approximately universal, so that increasing the circuit depth allows this family to approach Haar-like behavior.

The second set consists of diagonal, instantaneous quantum polynomial (IQP)-style ensembles~\cite{bremner2011classical, ni2013commuting, fujii2017commuting}. The base family, denoted here as $D_2$, is defined as
\begin{equation}
U
=
H^{\otimes n}
\exp\!\left(
i\sum_q \phi_q Z_q
+
i\!\!\sum_{(u,v)\in E}\!\! \phi_{uv} Z_u Z_v
\right)
H^{\otimes n},
\end{equation}

where the phases are sampled independently as $\phi_q,\phi_{uv}\sim\mathrm{Uniform}(0,2\pi)$, and the interaction graph $E$ is sampled from an Erd\H{o}s--R\'enyi ensemble $G(n,p)$ in which each edge is included independently with probability $p=d/(n-1)$. The mean degree $d$ is the tunable parameter of the family. Increasing $d$ introduces more $ZZ$ couplings and can make the output distribution appear increasingly random in a given measurement basis. However, this apparent randomness is basis dependent. The generators of $D_2$ commute and are diagonal in a fixed Pauli basis up to the surrounding basis change. As a result, $D_2$ can look expressive according to a single-basis output-statistics diagnostic while still retaining a strong underlying structure. The multi-basis ORS diagnostic is designed to reveal this limitation: a genuinely Haar-like ensemble should display Haar-like order statistics across measurement bases, not only in a privileged one.

To test how breaking this commuting structure affects expressivity and coverage, we also consider the noncommuting extensions $D_{2,XZ}$ and $D_{2,XZY}$. These are formed by stacking diagonal blocks in different Pauli bases: $D_{2,XZ}$ combines diagonal blocks in the $Z$ and $X$ bases, while $D_{2,XZY}$ adds a third block diagonal in the $Y$ basis. These constructions remain shallow and structured, but partially remove the basis dependence of $D_2$, providing an intermediate regime between purely commuting IQP-style reservoirs and fully random universal circuits.

\subsection{Tasks and datasets}
\label{sec:methods-tasks}

We evaluate the proposed diagnostics in both QELM and QRC settings, using synthetic tasks with controlled structure and real datasets with fixed, externally determined structure. The predictive performance is evaluated by the mean squared error (MSE) on a held-out test set or the $R^2$ Pearson coefficient.

\paragraph{Static QELM tasks.}
We consider two regression tasks. The first is a controlled random Fourier regression task \cite{xiong2024fundamental}. Given a scalar input $x\in[0,2\pi]$, the target function is
\begin{eqnarray}
    y(x)
=
\sum_{j=0}^{J_{\rm F}}
\left[
a_j \cos(jx) + b_j \sin(jx)
\right], \\ \nonumber
a_j,b_j\sim \mathrm{Uniform}(-1,1),
\label{eq:fourier-task}
\end{eqnarray}
with $J_{\rm F}=(3^{n_a}-1)/2$ and $n_a=6$ fixed throughout the sweep. The input is encoded using exponential angle encoding on $k_{\rm exp}=6$ qubits,
\begin{equation}
\ket{\psi(x)}
=
\bigotimes_{q=0}^{k_{\rm exp}-1}
R_z(-3^q x)H\ket{0},
\label{eq:exp-encoding}
\end{equation}
following Ref.~\cite{xiong2024fundamental}. This choice makes the target Fourier bandwidth compatible with the frequencies accessible to the encoding, so that limitations in performance can be attributed primarily to the reservoir feature map rather than to an insufficient input encoding. We use $n=6$ qubits, $5000$ samples in total, with $3500$ used for training and $1500$ for testing. The readout features are obtained from a fixed pool of $n_{\rm obs}=1000$ Pauli strings sampled uniformly at random and shared across all reservoir families and random seeds. In our tests, this pool size saturates $R_{\rm eff}$ for all families and avoids the structural bias introduced by restricting the observables to low-weight local Pauli operators.

The second QELM task is a quantum-chemistry regression problem. The goal is to predict the first excited electronic energy $E_1(R)$ of the LiH molecule from its corresponding electronic ground state $\ket{\psi_0(R)}$ at bond length $R$, using the dataset construction of Ref.~\cite{domingo2022optimal}. The molecular ground states are obtained by exact diagonalization of the electronic Hamiltonian for bond lengths $R_{\rm LiH}\in[0.5,3.5]$ a.u., and the LiH ground state is represented on $n=8$ qubits. The reservoir receives $\ket{\psi_0(R)}$ as input, while the target is $E_1(R)$. The dataset contains $300$ samples, split into training and test sets, with $30\%$ of the data used for testing. The test region is chosen as $R_{\rm LiH}\in[1.1,2.0]$ a.u., so that the model is evaluated on an interpolation/extrapolation regime not directly covered by the training samples. For this task, we measure all single-qubit Pauli observables $X$, $Y$, and $Z$, together with all two-local Pauli observables $\{O_iO_j\}, \ O_k \in \{X, Y, Z\} \ \forall k$.

\paragraph{Temporal QRC tasks.}
 We again consider one controlled synthetic benchmark and one real forecasting task. The synthetic temporal benchmark is the NARMA task, a standard nonlinear autoregressive problem that tests both nonlinearity and memory. For NARMA order $r$, the target sequence is generated as
\begin{equation}
\begin{aligned}
y(k)
&=
0.3\,y(k-1)
+
0.05\,y(k-1)\sum_{i=1}^{r} y(k-i)
\\
&\quad+
1.5\,u(k)u(k-r)
+
0.1 .
\end{aligned}
\label{eq:narma}
\end{equation}
We use $r=2$. The input signal $u(k)$ is encoded using the same exponential-encoding strategy, with order $n_a = 3$ and $k_{\rm exp}=3$. The QRC uses $n=6$ qubits: three input qubits, which are also included in the measured subsystem, and three hidden qubits that carry memory between time steps. We use the same train/test split as in the Fourier task, with $3500$ training points and $1500$ test points, and the same fixed pool of $n_{\rm obs}=1000$ random Pauli observables.

The real temporal benchmark is the EMSIG residential energy forecasting task~\cite{brucke2024benchmarking}. The data consist of energy measurements recorded by decentralized household energy management systems in the DACH region at $15$ minute resolution. We use EMS~3, corresponding to one building in the dataset, over the period from January 1, 2019 to December 31, 2020. The series is rescaled to $[0,1]$ with min--max scaling and resampled for one-hour-ahead prediction, reducing the strong short-lag autocorrelation present at the original $15$ minute resolution. We use $5000$ time points, with $30\%$ held out for testing. The scalar input is angle-encoded on one qubit. The QRC uses $n=4$ qubits, with one input qubit, two hidden memory qubits, and two measured qubits, where the measured subsystem includes the input qubit. Again, the observables are single and two-local Pauli observables.

For the random-gate families $G_1$, $G_2$, and $G_3$, we sweep the number of gates over $L\in\{10,20,50,100,200,1000\}$, while for the diagonal families $D_2$ and $D_{2,XZ}$ we sweep the mean interaction degree $d\in \{1,2,4,6,10\}$. Each reported diagnostic is averaged over $M=100$ reservoir instances.

\section{Results}
\label{sec:results}

We now evaluate the proposed diagnostics across reservoir families, noise regimes, and learning tasks. We first validate ORS as a scalable expressivity diagnostic by comparing it with standard expressivity measures. We then test its hardware-aware extension under simulated and experimental noise. Finally, we relate expressivity and coverage to predictive performance in both QELM and QRC settings.

\subsection{Order statistics, KL divergence, and dynamical complexity}
\label{sec:results-kl-ors}

We begin by comparing ORS with several established diagnostics of random-circuit complexity for the $G_3$ reservoir family at $n=6$ qubits, where the Hilbert-space dimension is still small enough for full-distribution and spectral diagnostics to be computed reliably. We sweep the gate count $L \in \{2,5,10,20,50,100,200\}$ and include a Haar reference. For each value of $L$, we sample $M=100$ independent reservoir instances. For every reservoir, we evaluate the ORS score gap $G_{\rm ORS}$ from Eq.~\ref{eq:ors-gap} for $K \in \{1,4,8\}$. We also compute the KL divergence from the Haar fidelity distribution, estimated from histograms of $N_{\mathrm{pairs}}=10^3$ pairwise fidelity samples with $n_{\mathrm{bins}}=75$, averaged over five independent repetitions to reduce sampling fluctuations. In addition, we report the ratio of consecutive level spacings $\overline{r}$ and the Krylov complexity, which probe the onset of quantum-chaotic spectral statistics and operator/state spreading, respectively \cite{domingo2024krylov}.
\begin{figure*}[t]
\centering
\input{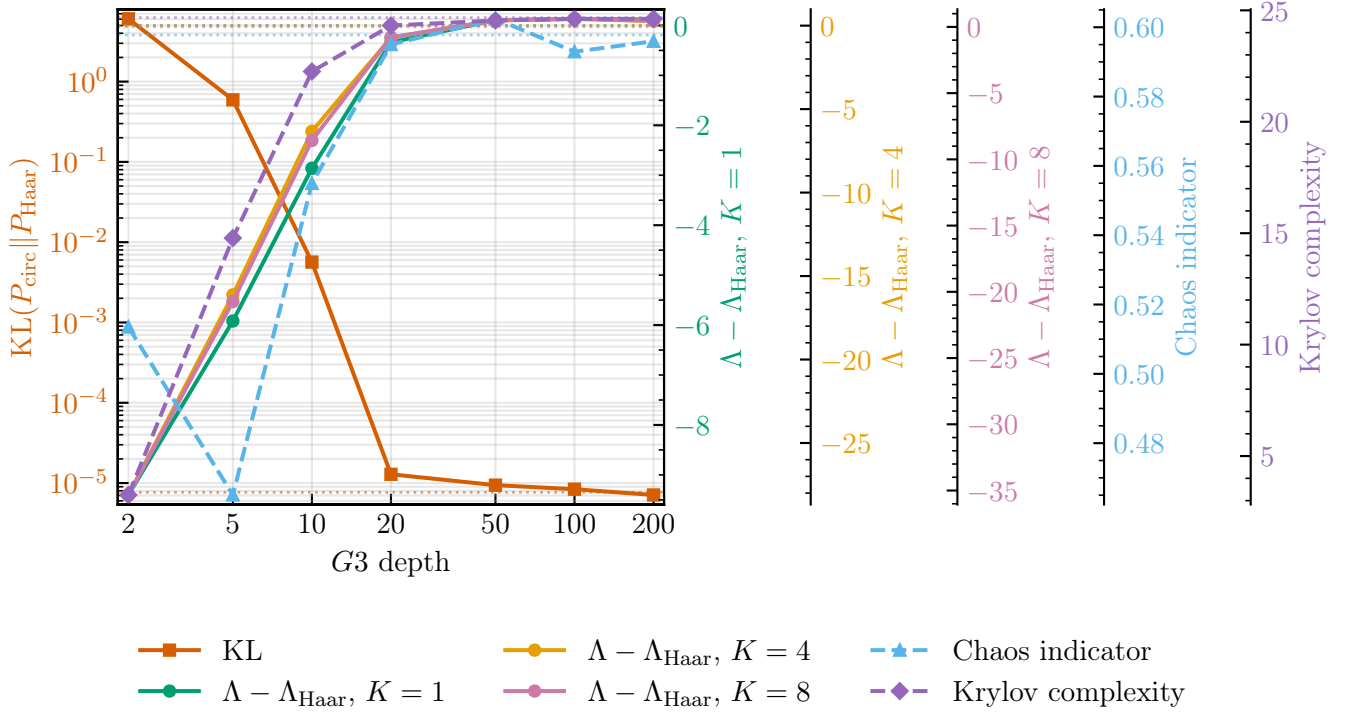}
\caption{Comparison of expressivity and dynamical-complexity diagnostics for the $G_3$ reservoir family at $n=6$. The KL divergence from the Haar fidelity distribution is shown in orange on the left logarithmic axis. The order-statistics scores $G_{\rm ORS}$ for $K \in \{1,4,8\}$ are shown on separate right axes, together with their Haar references. The dashed blue curve shows the chaos indicator based on the ratio of consecutive level spacings $\overline{r}$, and the dashed purple curve shows the Krylov complexity. All diagnostics identify the same transition from shallow structured circuits to Haar-like, chaotic behavior as the circuit depth increases.}
\label{fig:KL-ORS}
\end{figure*}
Figure~\ref{fig:KL-ORS} shows that the different diagnostics identify a consistent transition to Haar-like behavior. The KL divergence decreases by several orders of magnitude between $L=2$ and $L=20$, after which it saturates close to the finite-sampling histogram floor. Over the same range, the ORS scores for all tested values of $K$ approach their Haar references and remain stable at larger depths. This agreement indicates that the order statistics of only the largest output probabilities are already sufficient to detect the same expressivity transition captured by the full fidelity-distribution KL divergence. The spectral and dynamical diagnostics show the same qualitative transition. $\overline{r}$ approaches its chaotic reference as the depth increases, indicating the emergence of random-matrix-like spectral correlations. Similarly, the Krylov saturates near its Haar value once the circuit family has entered the highly scrambling regime. Thus, ORS is not only consistent with the KL expressivity diagnostic, but also with independent probes of quantum chaos and dynamical complexity.

The agreement across $K$ further shows that the detected transition is not too sensitive to the precise number of retained ranks, provided $K \ll D$. Increasing $K$ changes the numerical scale and variance of the ORS score, but not the depth at which the ensemble becomes Haar-like. We therefore use a fixed small value of $K = 4$ in the following experiments. Since the KL divergence, spectral statistics, and Krylov complexity become increasingly expensive or unreliable at larger system sizes, this small-system comparison validates ORS as the scalable expressivity diagnostic used throughout the rest of the work.

\subsection{Order statistics under noise}
\label{sec:results-noisy-ors}

We next test the hardware-aware ORS correction introduced in Sec.~\ref{sec:methods-noise}. Under the global depolarizing model, the corrected ORS gap $G_{\mathrm{ORS}}^{(f)}$ is invariant under the fidelity parameter $f$ for fixed reservoir instances and exact probabilities (see Eq.~\ref{eq:noisy-gap-invariance}). Table~\ref{tab:simulated-noise-invariance} reports a representative calculation at $f=0.5$. We verified numerically that the same corrected gaps are obtained for other values of $f$, up to numerical precision.
\begin{table}[t]
\centering
\begin{tabular}{c c c c}
\hline
$n$ & $G_1$ & $G_3$ & $D_2$ \\
\hline
6  & $-9.3$ & $+0.83$ & $-4.1$ \\
10 & $-290$ & $+0.72$ & $-76$ \\
12 & $-1.2\times 10^{3}$ & $-1.4$ & $-300$ \\
16 & $-2.5\times 10^{3}$ & $0.23$ & $-4.8\times 10^{3}$ \\
20 & $-3.2\times 10^{5}$ & $0.56$ & $-7.7\times 10^{4}$ \\
25 & $-1.0\times 10^{7}$ & $0.25$ & $-2.5\times 10^{6}$ \\
\hline
\end{tabular}
\caption{Representative noise-corrected ORS gaps under simulated global depolarizing noise. We report $G_{\mathrm{ORS}}^{(f)}=\overline{\Lambda}^{(f)}-\Lambda_{\mathrm{Haar}}^{(f)}$ for $K=4$ and $f=0.5$. The $G_1$ and $G_3$ reservoirs use $n \times 200$. The $D_2$ reservoir uses degree $d=10$, and the reported value is the average signed gap over the 50 random Pauli bases. Values close to zero indicate Haar-like behavior.}
\label{tab:simulated-noise-invariance}
\end{table}

The simulated results illustrate two points. First, the correction removes the trivial contraction of the output distribution toward the uniform distribution, allowing ORS gaps obtained at different effective fidelities to be compared on the same scale. Second, the corrected gap remains highly discriminative at larger system sizes. The universal $G_3$ family remains close to the Haar reference up to $n=25$, with residual deviations of order one. By contrast, the restricted $G_1$ family stays far from Haar-like behavior, with a deviation that grows rapidly with system size. The diagonal $D_2$ family is also far from Haar-like behavior when tested across several measurement bases, reflecting the basis dependence of its commuting structure. 

The IBM hardware experiment tests the same correction in a more realistic setting, where the effective fidelity is estimated from backend calibration data and the physical noise is not expected to be exactly global depolarizing. Table~\ref{tab:ibm-aachen-ors} reports ORS values measured on the IBM Aachen backend for selected $G_1$ and $G_3$ circuits. For each circuit, we estimate an effective fidelity $f_{\mathrm{mean}}$ from the calibrated one-qubit, two-qubit, and readout errors of the physical qubits used by the transpiled circuit (see Eq.~\ref{eq:block-error-decomposition}). Even under these hardware conditions, ORS separates shallow or structurally restricted circuits from deeper, more expressive ones. In particular, $G_1$ circuits remain far from the Haar reference, while sufficiently deep $G_3$ circuits move much closer to the corrected Haar baseline. This separation is visible across all tested system sizes.

\begin{table}[t]
\centering
\begin{tabular}{c c c c c}
\hline
$n$ & family & gates & $f_{\mathrm{mean}}$ & $G_{\rm ORS}^{(f)}$ \\
\hline
6  & $G_1$ & 50  & 0.93 & $-5.7$ \\
6  & $G_1$ & 300 & 0.94 & $-3.0$ \\
6  & $G_3$ & 50  & 0.94 & $-4.2$ \\
6  & $G_3$ & 300 & 0.94 & $+0.19$ \\
10 & $G_1$ & 50  & 0.88 & $-43$ \\
10 & $G_1$ & 300 & 0.89 & $-6.9$ \\
10 & $G_3$ & 50  & 0.89 & $-30$ \\
10 & $G_3$ & 300 & 0.90 & $-3.0$ \\
12 & $G_1$ & 50  & 0.86 & $-432$ \\
12 & $G_1$ & 500 & 0.86 & $-37$ \\
12 & $G_3$ & 50  & 0.88 & $-356$ \\
12 & $G_3$ & 500 & 0.88 & $-2.3$ \\
\hline
\end{tabular}
\caption{Hardware-aware ORS gaps measured on the IBM Aachen quantum backend. The effective fidelity $f_{\mathrm{mean}}$ is estimated from backend calibration data for the transpiled circuit and physical qubit block. The reported gap is $G_{\mathrm{ORS}}^{(f)}=\overline{\Lambda}^{(f)}-\Lambda_{\mathrm{Haar}}^{(f)}$, computed using the corresponding hardware-aware Haar baseline.}
\label{tab:ibm-aachen-ors}
\end{table}

Together, the simulated and hardware results show that the noise-corrected ORS gap provides a hardware-compatible expressivity diagnostic. In the ideal depolarizing setting, the correction makes ORS invariant to the effective fidelity, while on hardware it provides an effective way to compare reservoirs executed at different noise levels.

\subsection{Expressivity and coverage predict reservoir performance}
\label{sec:results-performance}

\begin{figure*}[t]
\centering
\resizebox{0.9\textwidth}{!}{%
    \input{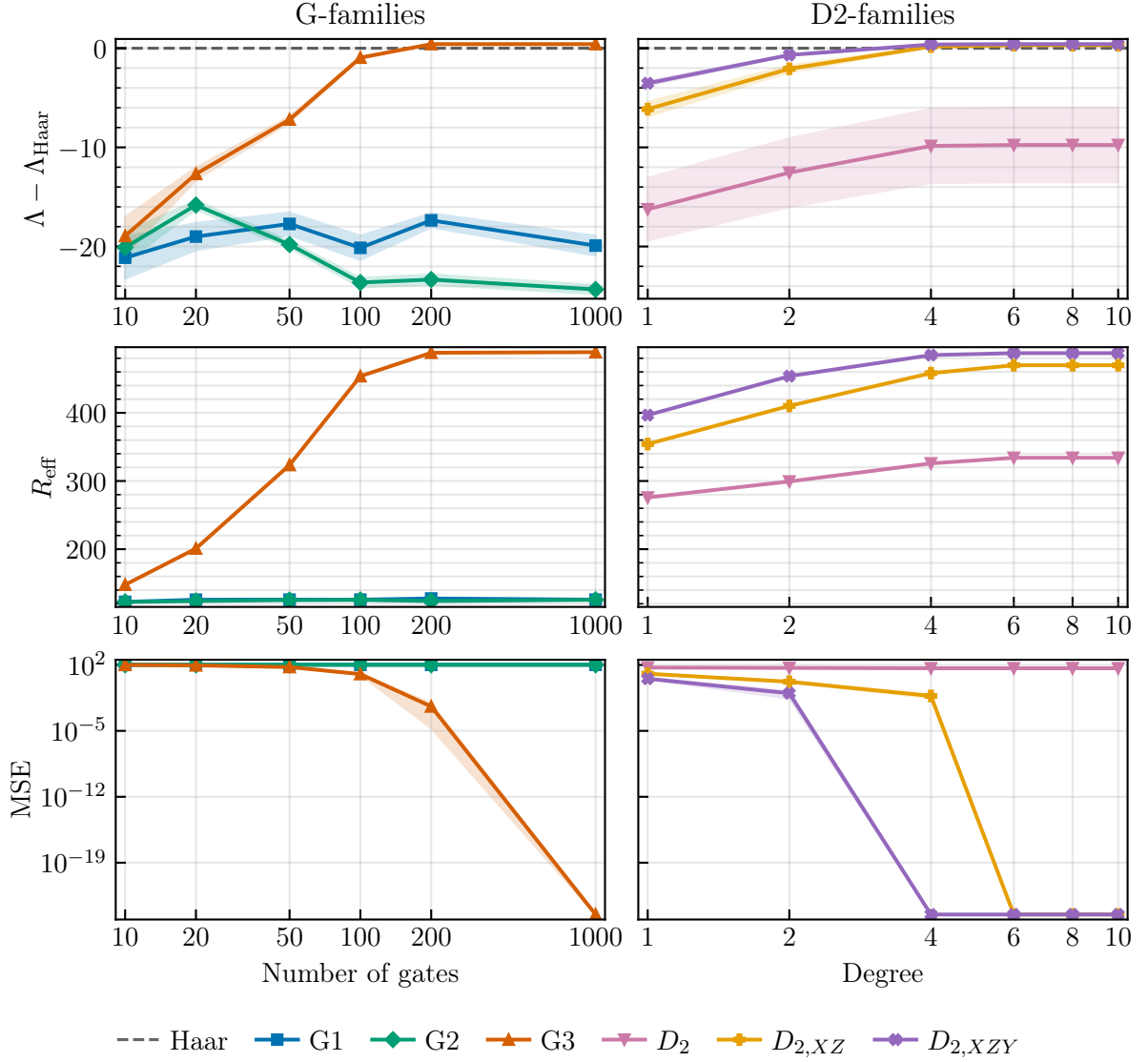}%
}
\caption{Quantum extreme learning machine results on the synthetic Fourier regression task for $n=6$. Top: ORS diagnostic, using the multi-basis diagnostic with $B=50$ random local Pauli bases for the diagonal families. Middle: effective rank $R_{\mathrm{eff}}$. Bottom: test Mean Squared Error (MSE).}
\label{fig:qelm-fourier-results}
\end{figure*}

\begin{figure*}[t]
\centering
\resizebox{0.9\textwidth}{!}{%
    \input{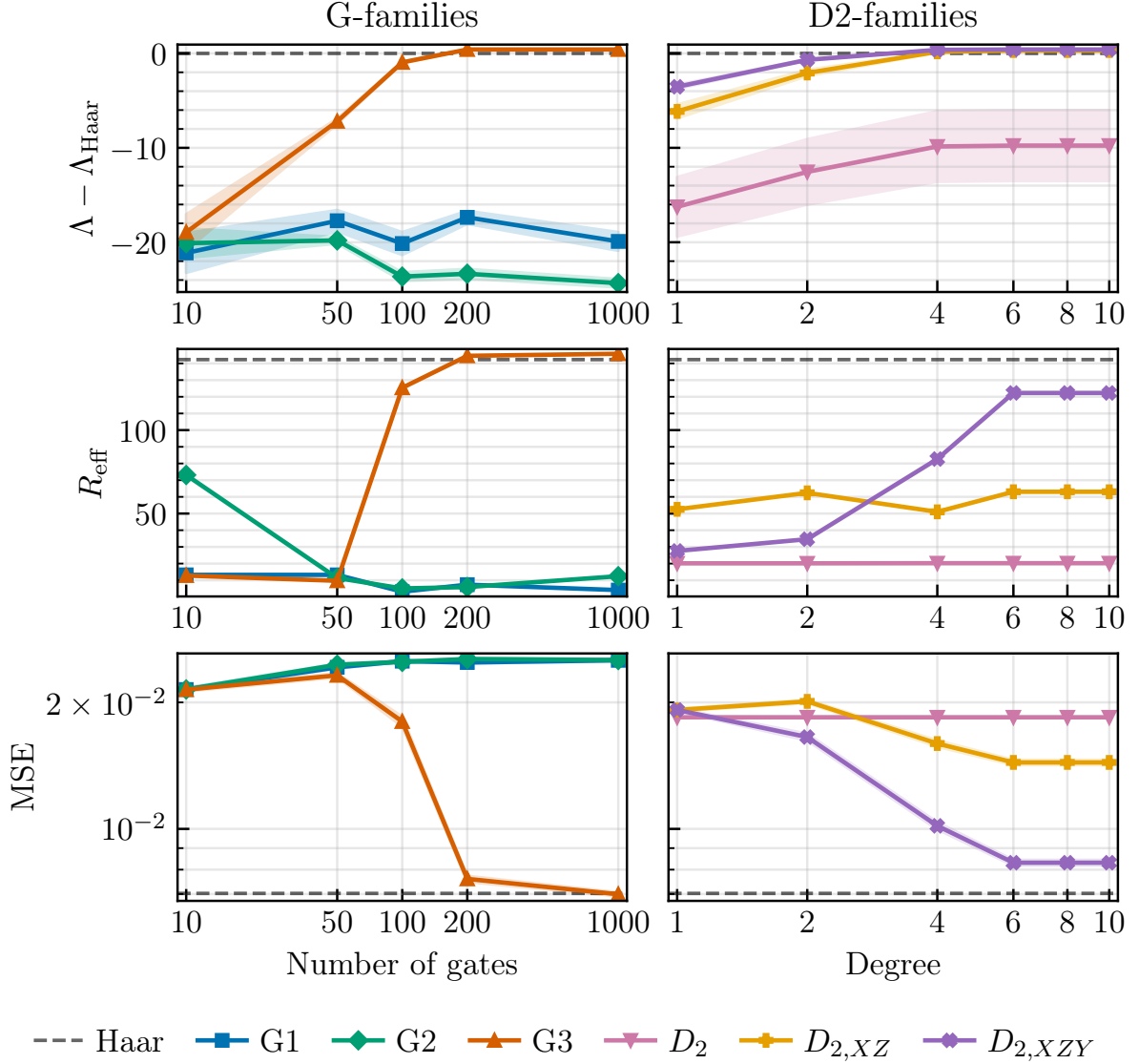}%
}
\caption{Same as Fig.~\ref{fig:qelm-fourier-results} for the quantum reservoir computing task on the synthetic NARMA dataset for $n=6$.}
\label{fig:qrc-narma-results}
\end{figure*}

We now relate the expressivity and coverage diagnostics to learning performance in the synthetic QELM and QRC tasks introduced in Sec.~\ref{sec:methods-tasks}. The results for the real-world datasets are presented in Appendix~\ref{app:real-data-results} and show qualitatively consistent behavior. In each figure, the top row reports the ORS-based expressivity diagnostic, the middle row shows the effective rank $R_{\mathrm{eff}}$, and the bottom row shows the test MSE.

Across both tasks, good performance occurs when expressivity and coverage are aligned. Reservoirs that move closer to the Haar reference in ORS also tend to produce feature matrices with larger effective rank, and these are precisely the regimes where the test error decreases. This is most evident for the universal random-gate family $G_3$: as the number of gates increases, its ORS gap approaches zero, $R_{\mathrm{eff}}$ rises, and the MSE drops. By contrast, the restricted families $G_1$ and $G_2$ remain far from the Haar reference, have lower effective rank, and perform poorly or deteriorate with depth.

The diagonal IQP-style families show the same general pattern, while also illustrating why ORS and $R_{\mathrm{eff}}$ are complementary. The commuting $D_2$ family remains clearly separated from its non-commuting extensions: it is further from Haar-like behavior across bases, has lower effective rank, and gives the weakest performance. Adding diagonal blocks in additional Pauli bases, as in $D_{2,XZ}$ and $D_{2,XZY}$, makes the output statistics more basis-robust and improves both coverage and prediction error.

The QRC results (Fig. \ref{fig:qrc-narma-results}), however, show that this correspondence is not exact. For $D_{2,XZ}$ and $D_{2,XZY}$, the multi-basis ORS diagnostic approaches the Haar reference at moderate degree, but $R_{\mathrm{eff}}$ remains below its maximum and the MSE does not reach the best achievable value. This indicates a partial decoupling between output-statistical expressivity and task coverage: a reservoir may look Haar-like in its ranked measurement probabilities while still failing to expose all the input-dependent directions needed by the readout.

Several effects could contribute to this decoupling. ORS uses only the top-$K$ ranked values of the largest probabilities, so it does not track which bitstrings carry those probabilities or the correlations between them. Residual structure in these correlations may still restrict the feature matrix even when the order statistics look Haar-like. Increasing the number of sampled bases or using larger values of $K$ could make ORS more sensitive to this structure. In addition, QRC performance depends on dynamical properties beyond static expressivity, especially memory capacity. Thus, multi-basis ORS should be understood as a scalable task-independent expressivity test, while $R_{\mathrm{eff}}$ reveals whether the chosen encoding, observables, and dynamics provide usable task-dependent features. Incorporating explicit memory-capacity diagnostics is a natural direction for future work.

\section{Conclusions}
\label{sec:conclusions}
We have introduced a two-axis framework for diagnosing QRs that is general across reservoir families, scalable, and compatible with noisy hardware. The first axis is ORS, a task-independent expressivity diagnostic that compares the largest output probabilities of a QR ensemble with an analytical Haar order-statistics baseline. Because it uses only the top-ranked outcomes rather than the full output distribution, the number of retained probabilities can be fixed independently of Hilbert-space dimension. Its closed-form depolarizing-noise correction further makes it directly applicable to hardware data. We also introduced a multi-basis extension, which tests whether Haar-like statistics persist across random local Pauli measurement bases, providing a stricter and more basis-robust expressivity diagnostic. The second axis is the effective rank $R_{\mathrm{eff}}$ of the feature matrix, a task-dependent coverage diagnostic that measures how much independent, input-dependent information is available to the linear readout.

Across synthetic and real QELM and QRC benchmarks, the two diagnostics jointly explain predictive performance. In the small-system regime, ORS agrees with KL divergence, spectral chaos indicators, and Krylov complexity, while remaining applicable under simulated depolarizing noise and execution on IBM quantum hardware. The multi-basis ORS score captures the intrinsic expressivity hierarchy of reservoir families, including the separation between commuting and non-commuting IQP-block constructions. The effective rank then explains when this expressivity becomes useful for learning: high-performing reservoirs are those whose expressive dynamics also generate a sufficiently rich feature matrix for the readout.

Several directions remain open. First, the multi-basis diagnostic raises the question of how many measurement bases are needed to reliably certify basis-robust expressivity, and how this number depends on system size and reservoir structure. Second, the framework is naturally suited to compare analog Hamiltonian reservoirs and gate-based reservoirs across different hardware platforms, since ORS is defined from measurement statistics rather than from a specific circuit model. Third, our noise treatment assumes an effective global depolarizing channel; extending the diagnostics to structured noise mechanisms, including coherent, correlated, and non-Markovian errors, will be important for making them predictive on near-term devices.

\section*{Author contributions}

L.D. and O.B. conceived the study and designed and carried out the numerical experiments. L.D. led the experimental analysis and writing of the manuscript. F.V. contributed to the writing of the manuscript and critically reviewed and revised the work. All authors discussed the results and reviewed the final manuscript.

\section*{Data and code availability}

The code used to generate the numerical results and figures presented in this work, together with the data required to reproduce the experiments, is available at \url{https://github.com/laiadc/order_statistics/}.

\section*{Acknowledgements}
O.B. is a fellow of Eurecat’s “Vicente López” Ph.D. grant program. This piece of research was carried out with the partial support of the following granted projects: SGR Grant No. 2021 SGR 01559 from the Generalitat de Catalunya, Departament de Recerca i Universitats; QMLCV Grant No. SDC007/25/000001 from the Generalitat de Catalunya, Departament d’Empresa i Treball; GRAIL Grant No. PID2021-126808OB-I00 and SUKIDI Grant No. PID2024-157778OB-I00 from the Spanish Ministry of Science and Innovation, the European Union, and the Agencia Estatal de Investigación; and Cátedra ENIA UAB-Cruïlla Grant No. TSI-100929-2023-2 from the Spanish Ministry of Economic Affairs and Digital Transformation and the European Union NextGenerationEU.

\addcontentsline{toc}{section}{References}
\bibliographystyle{apsrev4-2}
\bibliography{references}

\appendix

\section{Analytical Haar baseline for ORS}
\label{app:ors}

\begin{figure*}[t]
\centering
\resizebox{0.9\textwidth}{!}{%
\input{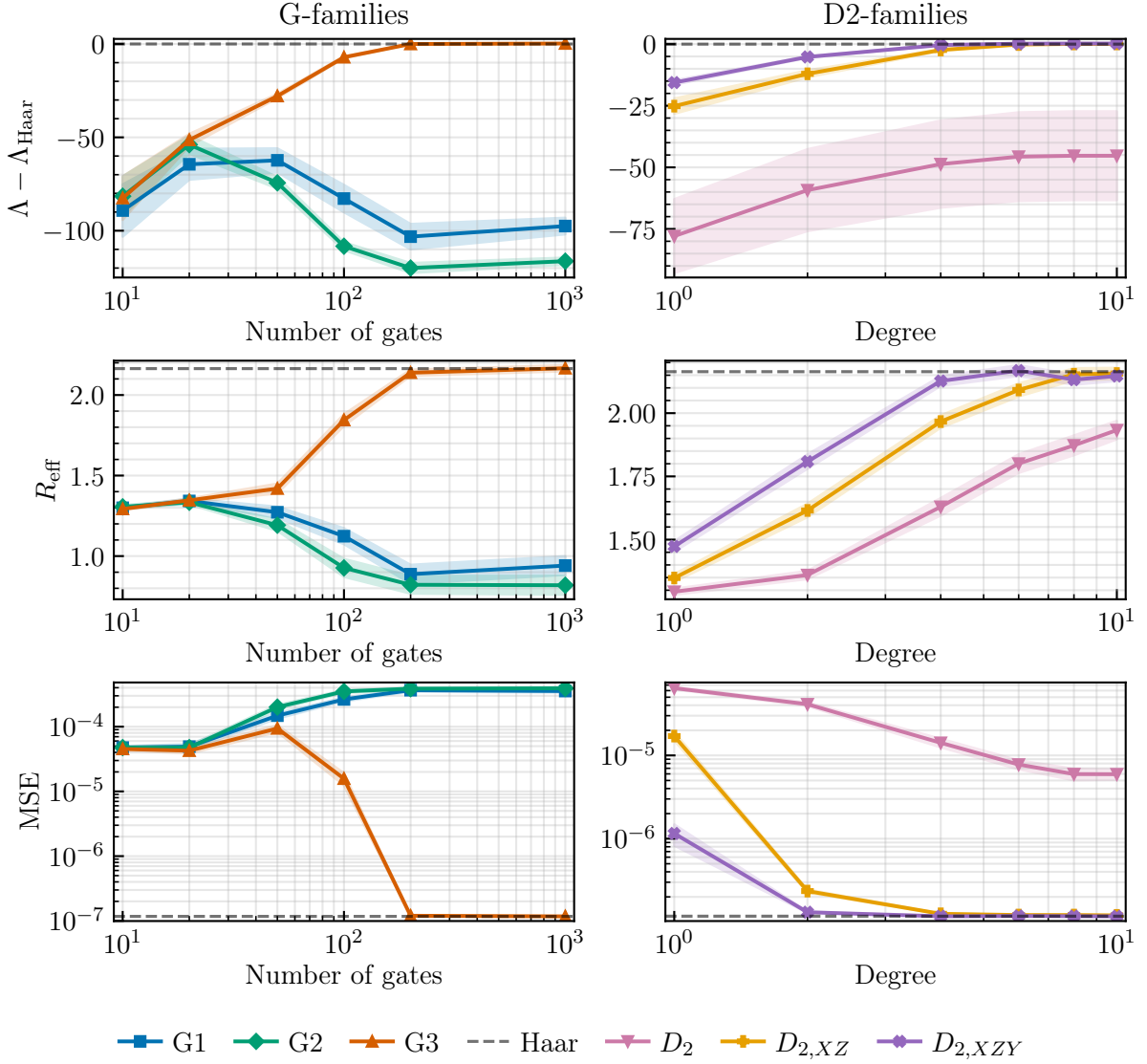}}
\caption{QELM results on the LiH excited-energy prediction task. The expressivity--coverage alignment observed in the synthetic Fourier task persists in the quantum-chemistry setting. Reservoirs that approach the Haar reference and develop larger effective rank achieve the lowest test MSE.}
\label{fig:lih-results}
\end{figure*}

\begin{figure*}[t]
\centering
\resizebox{0.9\textwidth}{!}{%
\input{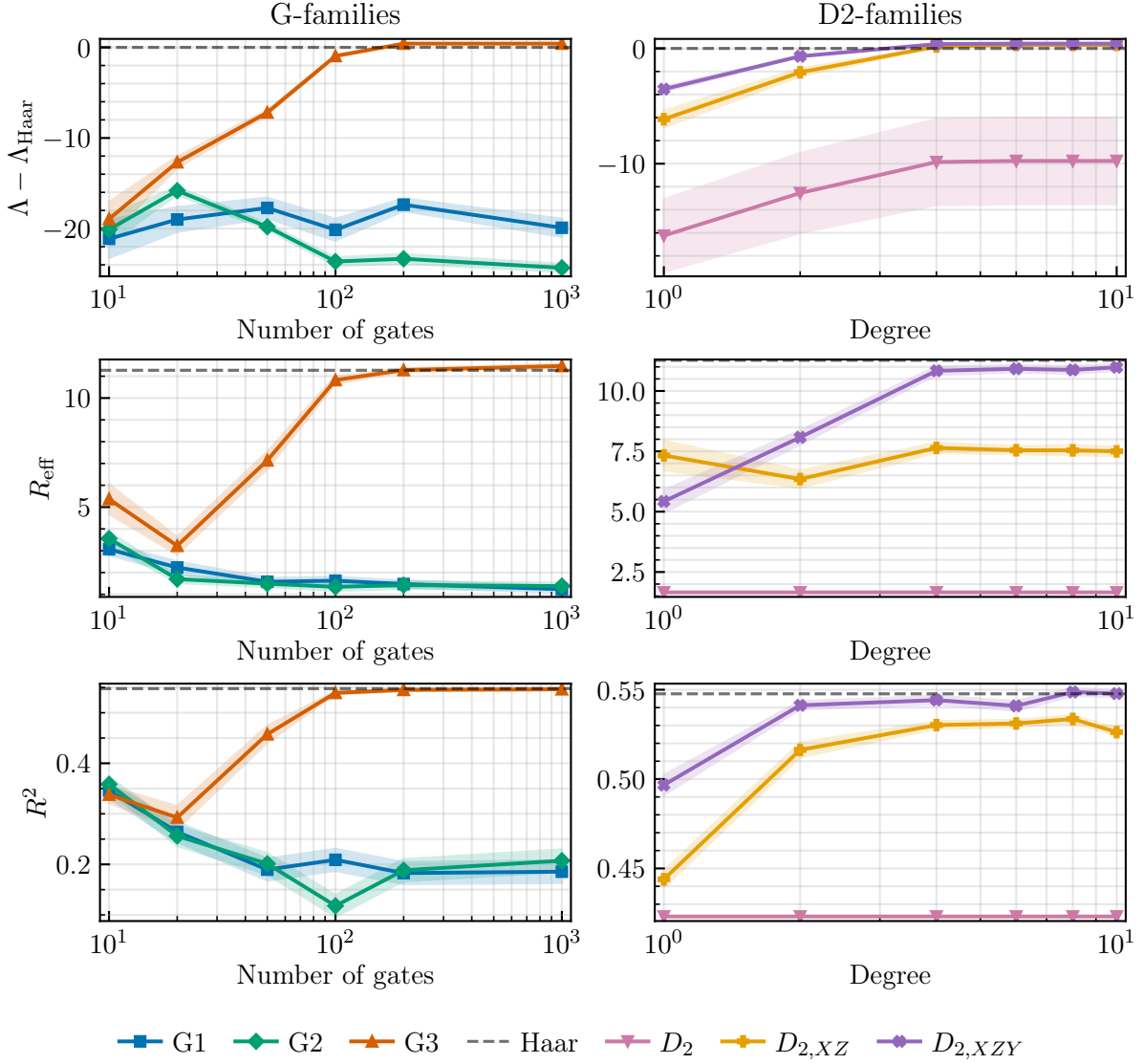}}
\caption{QRC results on the EMSIG residential energy forecasting task. The test metric is $R^2$, so higher values indicate better forecasting performance. The best-performing reservoirs are those for which strong expressivity diagnostics are accompanied by large effective rank.}
\label{fig:emsig-results}
\end{figure*}

The ORS diagnostic compares the log-likelihood of the observed top-ranked probabilities with its Haar expectation. This Haar reference can be evaluated analytically in the large-$D$ approximation, avoiding the need to sample Haar-random states. For a Haar-random state on $D=2^n$ basis states, let $p_{(k)}$ denote the $k$-th largest output probability. In the large-$D$ approximation, the order-statistics density of $p_{(k)}=x$ is
\begin{equation}
\begin{aligned}
P_k(x)
&=
\frac{a}{B(k,D-k+1)}
\exp(-k a x)
\\
&\quad \times
\left(1-\exp(-a x)\right)^{D-k},
\qquad
a=D-2 .
\end{aligned}
\label{eq:Pk-largeD}
\end{equation}
where $B(\cdot,\cdot)$ is the beta function. The contribution of rank $k$ to the Haar self-log-likelihood is
\begin{equation}
\Lambda_{k}^{\mathrm{Haar}}
=
\int P_k(x)\log P_k(x)\,dx .
\label{eq:lambda-k-haar-def}
\end{equation}
To compute this integral, introduce the change of variables
\begin{equation}
u=\exp(-a x).
\end{equation}
Under Eq.~\eqref{eq:Pk-largeD}, the transformed variable follows
\begin{equation}
u \sim \mathrm{Beta}(k,D-k+1).
\end{equation}
Moreover,
\begin{equation}
\begin{aligned}
\log P_k(x)
&=
\log a
-
\log B(k,D-k+1)
+
k\log u
\\
&\quad
+
(D-k)\log(1-u).
\end{aligned}
\end{equation}
Using the beta-distribution identities
\begin{equation}
\begin{aligned}
\mathbb{E}[\log u]
&=
\psi(k)-\psi(D+1),
\\
\mathbb{E}[\log(1-u)]
&=
\psi(D-k+1)-\psi(D+1).
\end{aligned}
\end{equation}

where $\psi$ is the digamma function, we obtain
\begin{align}
\Lambda_{k}^{\mathrm{Haar}}
&=
\log a
-
\log B(k,D-k+1)
\nonumber\\
&\quad + k\left[\psi(k)-\psi(D+1)\right]
\nonumber\\
&\quad + (D-k) \left[\psi(D-k+1)-\psi(D+1)\right].
\label{eq:lambda-k-haar}
\end{align}
Therefore, for the ORS score with harmonic weights $w_k=1/k$, the analytical Haar baseline is
\begin{equation}
\Lambda_{\mathrm{Haar},K}
=
\sum_{k=1}^{K}
\frac{1}{k}
\Lambda_{k}^{\mathrm{Haar}}.
\label{eq:lambda-haar-k}
\end{equation}

\section{Real-data reservoir benchmarks}
\label{app:real-data-results}

We also evaluate the expressivity--coverage diagnosis on two real-data benchmarks: the LiH quantum-chemistry regression task in the QELM setting and the EMSIG residential energy forecasting task in the QRC setting. These experiments provide an additional test of whether the relationship between expressivity, coverage, and performance persists when the data structure is fixed externally. As in the synthetic experiments, we use the multi-basis ORS diagnostic averaged over $B=50$ random local Pauli bases for the diagonal families.

The real-data benchmarks confirm the same qualitative picture as the synthetic experiments. In the LiH task, the $G_3$ family approaches the Haar-like regime at sufficient depth, develops a much larger effective rank than $G_1$ and $G_2$, and achieves the smallest test MSE. Among the diagonal families, the non-commuting constructions $D_{2,XZ}$ and $D_{2,XZY}$ are more expressive according to the multi-basis ORS diagnostic than the purely commuting $D_2$ family. This increased basis-robust expressivity is accompanied by a larger effective rank and lower prediction error. Thus, in this quantum-chemistry task, the multi-basis expressivity ranking is consistent with the observed performance ranking, while $R_{\mathrm{eff}}$ explains how this expressivity becomes available to the linear readout.

The EMSIG forecasting task gives the same conclusion in the QRC setting. The $G_3$ family again combines near-Haar ORS behavior, high effective rank, and the best forecasting accuracy. The restricted $G_1$ and $G_2$ families remain less expressive, have low effective rank, and perform substantially worse. Within the diagonal class, the multi-basis ORS diagnostic separates $D_2$, $D_{2,XZ}$, and $D_{2,XZY}$, matching the hierarchy observed in both $R_{\mathrm{eff}}$ and test $R^2$. The non-commuting variants, particularly $D_{2,XZY}$, produce more basis-robust output statistics, expose more independent feature directions, and achieve better forecasting performance than the purely commuting $D_2$ family.

\end{document}